\documentclass[10pt,conference]{IEEEtran}
\IEEEoverridecommandlockouts
\usepackage{cite}
\usepackage{amsmath,amssymb,amsfonts}
\usepackage{algorithmic}
\usepackage{graphicx}
\usepackage{textcomp}
\usepackage{xcolor}
\usepackage{hyperref}
\def\BibTeX{{\rm B\kern-.05em{\sc i\kern-.025em b}\kern-.08em
    T\kern-.1667em\lower.7ex\hbox{E}\kern-.125emX}}
\begin{document}

\bibliographystyle{IEEEtran}

\title{Towards security defect prediction with AI
\thanks{Funding blinded for review.}
}

\author{\IEEEauthorblockN{Carson D. Sestili}
\IEEEauthorblockA{\textit{CERT - Software Engineering Institute} \\
\textit{Carnegie Mellon Univerity}\\
Pittsburgh, USA \\
sestilic@gmail.com}
\and
\IEEEauthorblockN{William S. Snavely}
\IEEEauthorblockA{\textit{CERT - Software Engineering Institute} \\
\textit{Carnegie Mellon Univerity}\\
Pittsburgh, USA \\
will.snavely@gmail.com}
\and
\IEEEauthorblockN{Nathan M. VanHoudnos}
\IEEEauthorblockA{\textit{CERT - Software Engineering Institute} \\
\textit{Carnegie Mellon Univerity}\\
Pittsburgh, USA \\
nmvanhoudnos@cert.org}
}

\maketitle

\begin{abstract} 
In this study, we investigate the limits of the current state of the
art AI system for detecting buffer overflows and compare it with
current static analysis tools. To do so, we developed a code
generator, {\em s-bAbI}, capable of producing an arbitrarily large number of
code samples of controlled complexity. We found that the static
analysis engines we examined have good precision, but poor recall on
this dataset,
except for a sound static analyzer that has good precision and recall. We
found that the state of the art AI system, a memory network modeled
after Choi et al. \cite{choi2017}, can achieve similar performance to
the static analysis engines, but requires an exhaustive amount of
training data in order to do so.
Our work points towards future approaches that may solve these
  problems; namely, using representations of code that can capture
  appropriate scope information and using deep learning methods that
  are able to perform arithmetic operations.
\end{abstract}


\section{Introduction}

Predicting security defects in source code is of significant national security interest.
It is ideal to detect security defects during development,
before the code is ever run to expose those defects.  The current best
methods to find security defects before running code are static
analysis tools, a variety of which exist and model software in
different ways that are all useful for different kinds of flaws.
Developers of static analyzers carefully equip them with rules about
program behavior, which are used to reason about the safety of the
program if it were to run.

However, static analyzers are known to be insufficient at finding
flaws.  The Juliet Test Suite
\cite{Boland2012-ow,Black2018-ae,Black2018-le} is a collection of
synthetic code containing intentional security defects across hundreds
of vulnerabilities in the Common Weakness Enumeration standard,
labeled at the line-of-code level.  Even state-of-the art
static analyzers perform poorly at finding the defects in Juliet,
issuing too many false positives and also too many false negatives
\cite{Wagner2005-pp,Emanuelsson2008-hu,Johnson2013-lc,Goseva-Popstojanova2015-jg}.
Finding an automated method that detects security defects in Juliet,
with acceptably low false-positive and false-negative counts, is an
open problem \cite{Black2018-le}.

Artificial intelligence (AI) as a broad field, and machine learning as
a tool for AI, have both seen great advances over the last decade.
An attractive property of many data-driven AI systems that ingest large
amounts of data and learn an accurate model of the underlying
distribution is that the features that are most important to describe
the model do not need to be explicitly extracted.  These systems often
discover ``hidden'' features that a human hand-crafting a model might
never think to include.  For recent problems where features are hard
to describe but data is abundant, deep learning has proven to be an
especially useful technique, both for extracting relevant features and
for creating predictive models that use them.
For example, deep learning methods are used in the state of the art
solutions for image recognition, surpassing human performance
\cite{He2015};
current question answering systems are able to rival 
human performance for a subset of tasks focusing on answering
realistic reading comprehension questions \cite{squad-zk,Yu2018-bi}; and
a language modeling technique that represents words from natural-language
corpora in two different languages, generating similar representations
for words with similar semantic content, leads to improved machine
translation \cite{Zou2013}.

To benchmark the advances of AI, it has been useful to generate datasets
that challenge the current state of the art. The bAbI dataset \cite{weston2015}
is a set of simple English-language stories and accompanying reading
comprehension questions and correct answers. It establishes a target
for question-answering AI systems, such that any such system must
be able to achieve high performance on the bAbI dataset to be
considered successful. This challenge successfully improved the state
of question-answering AI, prompting the development of better
machine learning models that achieved greater performance on
the bAbI dataset, see \cite{bAbI:URL} for a comprehensive list.

This environment naturally leads to investigating whether data-driven
techniques can succeed in modeling existing source code
datasets.  Some work has already been done in AI on code
\cite{Allamanis2017-wv}, mostly on problems about modeling gaps in the
code from surrounding context; for example, the model of code2vec
\cite{Alon2018-mv} is able to suggest the name of a method by reading
the code in the method body.  However, there is little work done yet
in using AI to model security defects.  The most significant barrier
to this work is the lack of large enough quantities of data with
accurately labeled security defects on which to train machine learning
models.  For example, the Juliet test suite is not ideal for training
such models, even though it has labeled defects, because it has too
few examples and is too complex.

Choi et al \cite{choi2017} are to our knowledge the first to propose a
data-driven model to find security defects in code at the line-of-code
level. They follow an approach similar to bAbI \cite{weston2015} and
show that a deep learning model can in principle be used to model
buffer overflows.  They construct a synthetic code dataset with
labeled safe and unsafe buffer writes, which we call {\em CJOC-bAbI},
and training and testing on this codebase.  However, their dataset
does not contain real code--it is not syntactically valid C, and does
not compile.  Moreover, none of the functions in their dataset have
any non-trivial control flow--there are no conditionals or loops,
which would always be found in code developed for real applications.
Although their work is a significant first step in demonstrating the
utility of deep learning for code modeling, these weaknesses leave
important questions unanswered.

To more thoroughly test the utility of AI for code understanding, a
better dataset is needed, that can demonstrate whether a machine
learning system can accurately model non-trivial control flow.  If
this dataset is free of other numerous sources of complexity usually
found in source code, then researchers will be able to cleanly focus
on developing models that are able to capture the relevant details of
control flow.

The contributions of this work are as follows:
\begin{itemize}
\item We produce a code dataset that we call {\em s-bAbI}, containing
  syntactically-valid C programs with non-trivial control flow, and
  safe and unsafe buffer writes labeled at the line-of-code
  level. Although this code is simple as compared to real-world code,
  static analyzers do not perform as well as expected.
\item We demonstrate the limits of the work by Choi et al. \cite{choi2017}, showing
  that their deep learning model needs too much training data to be
  reasonably used to model security defects, and that their approach
  necessarily overfits to synthetic datasets, not easily generalizing
  to real code.
\item We point towards future approaches that may solve these
  problems; namely, using representations of code that can capture
  appropriate scope information and using deep learning methods that
  are able to perform arithmetic operations.
\end{itemize}

\section{Motivating Example}

\begin{figure}[h]
\includegraphics[width=8cm]{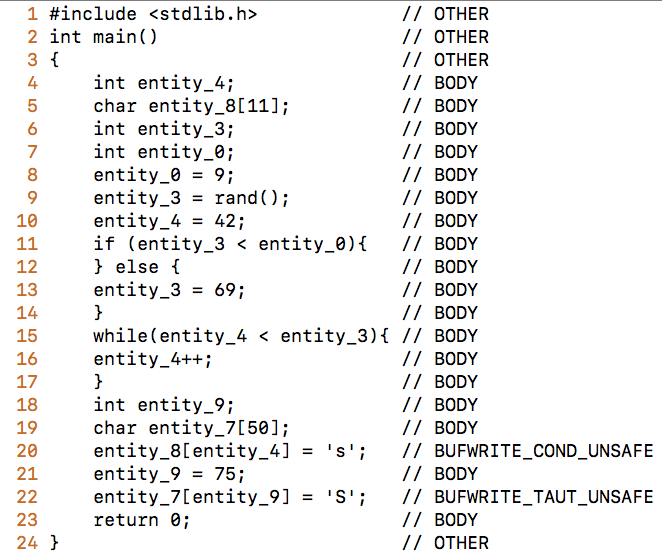}
\caption{A sample function from our generated dataset.}
\label{fig_sbabi_example}
\end{figure}

An example file from our dataset (\verb|77056b8250.c|) with non-trivial control flow is shown in Figure \ref{fig_sbabi_example}.

It is fairly easy to tell that the buffer write at line 22 is unsafe.
Doing so requires resolving the value of the index variable \verb|entity_9| and the length of the array \verb|entity_7|, and observing that the index is greater than the length of the array.
No knowledge of the control flow structure of the code granted by the \verb|if| statement and the \verb|while| loop is necessary to decide that this buffer write is unsafe.
The existing CJOC-bAbI dataset frames all of its buffer overflow examples in a setting not much more complicated than this example, only requiring variable lookup and a single comparison.
We would expect that any reasonable AI system should be able to recognize that this line is unsafe.

It is more challenging to tell that the buffer write at line 20 is unsafe.
To see this, first we note that \verb|entity_3| can take any integer value on line 9.
This means that the condition in the \verb|if| statement on line 11 is sometimes true and sometimes false.
Whenever it is false, the \verb|else| branch executes, setting \verb|entity_3| to 69.
Since \verb|entity_4| is originally 42 (from line 10), which is less than 69, the \verb|while| loop on line 15 executes, setting the value of \verb|entity_4| to 69.
But the length of the array \verb|entity_8| is 11 (from line 5), so the index is greater than the length of the array at the buffer write on line 20, and this access is unsafe.
We would expect that an AI system would have more trouble with the harder task of deciding that this line is unsafe.

To our knowledge, ours is the first paper showing that a deep learning model has the ability to predict security defects, at the line-of-code level, in synthetic code with control flow elements like conditionals and loops.

\section{Approach}
We develop a synthetic code generator that addresses shortcomings in both the CJOC-bAbI dataset and the Juliet Test Suite that prevent these datasets from being useful for developing a data-driven approach to security defect detection.
The CJOC-bAbI dataset is not made of syntactically valid C code, and none of the examples contain code with any non-trivial control flow elements like conditionals, loops, or variables with unknown values, so a data-driven technique trained and tested on this dataset provides a weak argument that a similar method will work on real code.
On the other hand, the Juliet Test Suite dataset is far too small and too complex for current data-driven methods to learn to predict the labeled security defects.
Our dataset presents a compromise, as it is complex enough to contain non-trivial control flow elements, but simple and large enough to successfully train data-driven approaches.
In this section, we describe details of our dataset, give an overview of the static analyzers we test on this dataset, and describe the deep learning architecture we compare against these static analyzers by training and testing on this dataset.

\subsection{s-bAbI: AI tasks for source code}

We propose a synthetic C source code dataset generator, which we call {\em s-bAbI}.
The purpose of this generator is to provide the simplest possible source code, labeled at the line-of-code level with safe and unsafe buffer writes, that contains control flow structures found in real-world code.
The motivation is analogous to that of the original English-language bAbI dataset, but adapted to the purpose of code understanding, namely, that any automated system that claims to be able to detect buffer overflows at the line-of-code level should achieve high predictive performance on this dataset.

\subsubsection{Code}

Each file in the s-bAbI dataset is syntactically-correct C source code, consisting of a single \verb|void main()| function.
This is an improvement over the dataset developed by Choi et al \cite{choi2017}, which contains syntactically-invalid ``C-like'' code examples.

Every s-bAbI code example has at least one of the following three elements that are found in any non-trivial real-world code:
\begin{itemize}
\item Conditional statements (\verb|if|/\verb|else|).
\item Loops (\verb|for| and \verb|while|).
\item Variables with unknown values (set from \verb|rand()|).
\end{itemize}

We acknowledge that these are not the only constructions that are found in non-trivial real-world code, e.g. function calls would be the next reasonable step to add complexity.
However, even these three elements add enough complexity to identify
weaknesses of both existing static analyzers and the deep learning
model proposed by Choi et al. \cite{choi2017}, meeting the purpose of this dataset.

Every line of an s-bAbI instance that does not modify control flow contains exactly one of the following:
\begin{itemize}
\item Integer variable declaration
\item Integer variable set to an integer literal
\item Integer variable set to the result of \verb|rand()|
\item Integer variable increment via \verb|++|
\item Character array declaration with integer literal length
\item Character array set at integer variable index ({\em buffer write})
\end{itemize}
Each buffer write is either {\em safe} or {\em unsafe}, meaning that the character array is written to at a value less-than-or-equal-to its length, or greater than its length, respectively.

\subsubsection{Labels}\label{sec_labels}
The motivation behind our labeling scheme is that it is ideal for tools to be able to identify {\em all} potentially unsafe lines in a function, not only those which can provably be reached, so that the developer can be alerted of all bugs simultaneously.

The generator that produces the s-bAbI dataset produces exactly one label per line, from the following six labels:
\begin{itemize}
\item \verb|BUFWRITE_COND_SAFE|
\item \verb|BUFWRITE_COND_UNSAFE|
\item \verb|BUFWRITE_TAUT_SAFE|
\item \verb|BUFWRITE_TAUT_UNSAFE|
\item \verb|BODY|
\item \verb|OTHER|
\end{itemize}
 
The {\em conditional} labels \verb|BUFWRITE_COND_SAFE|, \verb|BUFWRITE_COND_UNSAFE| refer to lines that contain exactly one buffer write, which is provably either safe or unsafe (buffer overflow), respectively, such that reasoning about the control flow is required to prove whether the write is safe or unsafe.
We refer to this kind of vulnerability as ``conditional'' since the safety of these buffer writes conditionally depends on the program's control flow.
For example, an integer index may be set to a value less than the array's length in the true branch of an \verb|if| statement, and a value greater than or equal to the array's length in the false branch.
Knowing whether the buffer write afterward is safe requires knowing which branch was taken.
For instances where an integer index is first set to an unknown value through the use of \verb|rand()|, the label is \verb|BUFWRITE_COND_SAFE| if {\em every} initial value of this variable results in safe control flow, and \verb|BUFWRITE_COND_UNSAFE| otherwise.

The {\em tautological} labels \verb|BUFWRITE_TAUT_SAFE|, \verb|BUFWRITE_TAUT_UNSAFE| refer to lines that contain exactly one buffer write, which is again provably either safe or unsafe, respectively, but whose safety can be determined without reasoning about control flow.
We call this kind of vulnerability ``tautological'', abusing this term slightly, because their safety depends on no information about the program's control flow.
In these instances, the integer index is always set exactly once in the main scope of the function, not inside any of the control flow structures.

The \verb|*COND*| buffer writes always occur in the main scope of the function, and are always reachable.
The \verb|*TAUT*| labels are allowed to occur inside the scope of both reachable and unreachable control flow structures, so they denote the safety of these lines under the assumption that they can be reached.
Since multiple buffer writes can occur in a file, they are labeled as if the program would reach these lines even in the case of an earlier unsafe write (that is, as if the program would not crash after trying to unsafely write to a buffer).

The \verb|BODY| label refers to a line in the body of the function that does not contain a buffer write.
The \verb|OTHER| label refers to a line outside the body of the function.

\subsection{Static Analysis Engines}

A variety of static analysis techniques are applicable to the buffer 
overflow detection problem. Consider our motivating example, in which
we wish to determine, for a given array index operation in a 
program, if it is possible for the index to be outside the 
bounds of the array. In the general case, for a Turing-complete
programming language, answering this question is undecidable.
Analyses therefore make trade-offs between soundness and completeness
\cite {chess2004static}. A sound analysis will never accept an
incorrect program---it yields no false negatives. A complete analysis
will always accept a correct program---it yields no false positives.
Due to the noted theoretical limitations, a sound program analysis
cannot be complete, nor a complete analysis sound.

We consider the standard statistical metrics {\em precision} and {\em recall},
defined by $\text{precision} = \frac{\text{TP}}{\text{TP} + \text{FP}}$ and
$\text{recall} = \frac{\text{TP}}{\text{TP} + \text{FN}}$, as measures of
soundness and completeness. Since a sound program yields no false negatives,
recall = 1, and since a complete program yields no false positives,
precision = 1.

The theory of abstract interpretation provides a framework for
sound static analyses \cite{cousot1977abstract}. Under this approach,
one develops sound abstractions of program semantics, which
are used to compute over-approximations of program behavior. For
example, for each array access in a program, we could over-approximate 
the domains of the array size and index, and soundly identify
bounds violations---at the risk of yielding false positives,
decreasing precision.

Other analysis approaches sacrifice soundness with the goal 
of yielding fewer false positives and/or improving analysis 
performance. Loop analysis is one domain where this trade-off
can be made. For example, a static analysis might only 
compute the effects of a finite number of loop iterations,
instead of soundly computing the effect of the loop. For the
array bounds problem, this heuristic certainly could yield
false negatives, for instance if a loop contains an array
index that is initially safe, but which goes out of bounds
after some number of iterations. Saxena discusses loop
analysis in more detail \cite{saxena2009loop}.

A 2004 study of static analysis tools applied to buffer
overflows provides an overview of techniques that
remains relevant, touching on abstract interpretation, symbolic
execution, and model checking \cite{Zitser2004-ep}.
This study also discusses some of the practical problems
tools face, such as aliasing and inter-procedural analysis.
A large body of research has focused on refining and extending
these techniques, for example improving the precision and
efficiency of sound analyzers (\cite{logozzo2008pentagons,nazare2014validation,cousot2009does,payet2018checking}),
and building better symbolic execution engines \cite{baldoni2018survey}.

Our work uses four static analysis tools for C. Three of these tools
are open source: Frama-C \cite{kirchner2015frama}, the Clang static
analyzer \cite{Clang:URL}, and Cppcheck \cite{Cppcheck:URL}.
We also used a commercial static analysis tool (anonymized).
Frama-C provides ``a collection of scalable, interoperable, and sound
software analyses'' for ISO C99 source code \cite{kirchner2015frama}.
We used the value analysis plugin to Frama-C to look for buffer overflows.
This plugin uses abstract interpretation to compute information about
integers and pointers in C programs, and issues warnings about possible
out-of-bounds accesses. See Kirchner et al. \cite{kirchner2015frama} for more information 
about the abstract domains employed by Frama-C to model these program entities.
The Clang static analyzer is based on symbolic execution,
and, by default, makes use of unsound heuristics such as loop unrolling to
contend with state space explosion\cite{Coughlin-2017}. 
We believe Cppcheck also makes use of unsound heuristics, though little has been published 
about the specific approach of this tool. The commercial tool we used is well-known 
to be unsound.

\subsection{Memory Network}

We test the effectiveness of a data-driven, deep learning-enabled approach to predicting buffer overflows using a {\em memory network} architecture.
Weston et al \cite{weston2014} and Sukhbaatar et al. \cite{sukhbaatar2015} developed and refined memory networks to answer questions about English-language stories in the bAbI dataset, as described in greater detail in Section \ref{sec_related_work}.
Choi et al. \cite{choi2017} performed the first application of memory networks to the task of buffer overflow prediction in synthetic code.
We use the same architecture as Choi et al. \cite{choi2017} , except for five key differences, noted below with asterisks.

\subsubsection{Data Preprocessing}
We prepare a C file for processing by the following procedure.
\begin{enumerate}
\item Split the file into lines of code, and tokens within each line
  of code, using the \verb|libclang| interface to \verb|clang| \cite{Clang:URL}.
\item (*) Prepend each line with a special token \verb|<line |$i$\verb|>| indicating its line number. Whereas Choi et al. \cite{choi2017} do not add line numbers, we do, as a simple preprocessing step that does not require any additional parameters to be optimized in the model. Line numbers provide sequential information to the memory network, which has no way of representing line order.
\item Using a consistent mapping from tokens to integers $\{1, 2, \dots, V - 1\}$ where $V - 1$ is the number of unique tokens, convert each token to an integer.
\item Save the integer tokens in an array, using zero padding on the right (dimension 1, to fill the rest of each line) and bottom (dimension 0, to fill the missing lines), obtaining an array of shape $N_F \times N \times J$, where $N_F$ is the number of files, $N$ is the maximum number of lines in a file, and $J$ is the maximum number of tokens in a line. The elements of the array are the $V$ integers $\{0, \dots, V-1\}$.
\end{enumerate}

\subsubsection{Network Architecture}
We describe the forward pass of the memory network.

{\bf Input:}
\begin{itemize}
\item A program code $X$ $[N \times J]$, consisting of $N$ lines $X_1, \dots, X_N$, where each line $X_i$ is a list of integer tokens $w_i^1, \dots, w_i^J$
\item A query line $q$ $[1 \times J]$, equal to one of the lines $X_i$ encoding a buffer write
\end{itemize}

{\bf Embedding:}
We fix an embedding dimension $d$ and establish two learnable embedding matrices $E_\text{val}$ and $E_\text{addr}$, both of dimension $V \times d$.
Letting $A$ represent both $E_\text{val}$ and $E_\text{addr}$, we encode each integer token twice, letting $Aw_i^j$ $[1 \times d]$ be the $w_i^j$-th row of $A$.
We use the position encoding of Sukhbaatar et al. \cite{sukhbaatar2015} to encode the position of tokens within each line: for $i = 1, \dots, N$, define $m_i$ $[1 \times d]$ by
$$
m_i = \sum_{j = 1}^J l_j \cdot Aw_i^j
$$
where $\cdot$ denotes elementwise multiplication and $l_j$ $[1 \times d]$ is defined by its $k$-th element as
$$
l_j^k = (1 - j / J) - (k / d) (1 - 2 j / J)
$$
(*) We apply Dropout \cite{Srivastava2014-qr, Gal2016-bb} with parameter 0.3 to each line, so that
$$
m_i \hspace {5pt} [1 \times d] = \text{Dropout}_{0.3}(m_i)
$$
We store the lines $m_i$ encoded by $E_\text{val}$ in a matrix $M_\text{val}$ $[N \times d]$, and store the lines encoded by $E_\text{addr}$ in a matrix $M_\text{addr}$.
We embed the query line $q$ by $E_\text{addr}$ and store the result in $u^1$ $[1 \times d]$.

{\bf Memory search:}
For each ``hop number'' $h = 1, \dots, H$ in a fixed number of ``hops'' $H$:
\begin{align*}
p \hspace{5pt} [N \times 1] & = \text{softmax}(M_\text{addr}u^T) \\
o \hspace{5pt} [1 \times d] & = \sum_{i = 1}^N p_i (M_\text{val})_i \\
(*) \hspace{5pt} r \hspace {5pt} [1 \times d] & = R_h o \\
(*) \hspace{5pt} s \hspace {5pt} [1 \times d] & = \text{Norm}_h(r) \\
u^{h+1} \hspace {5pt} [1 \times d] & = u^h + s
\end{align*}
where $R_h$ $[d \times d]$ is an internal learnable weight matrix, and $\text{Norm}_h$ is a batch normalization layer \cite{Ioffe2015-fj} with parameters $\theta_h$.

{\bf Classification:}
$$
\widehat{y} \hspace {5pt} [2 \times 1] = \text{softmax}(W(u^H)^T)
$$
where $W$ $[2 \times d]$ is a learnable weight matrix.

The forward pass of the network is effectively an iterative inner-product search \cite{Ram2012} matching the current query line $u^h$, which changes with each processing hop, against each line $m_i$ of the stored memory, which remains fixed.

We use the standard cross-entropy loss function to evaluate goodness of fit.
To train, we use gradient descent to minimize the loss across the training set as a function of the learnable parameters $E_\text{val}, E_\text{addr}, R_h, \theta_h, W$.
Note that the cross-entropy loss penalizes false negatives and false positives equally.
Our optimization attempts to maximize the combination of precision and recall, or viewed differently, attempts to approximate both soundness and completeness.

Like Choi et al. \cite{choi2017}, we use the Adam \cite{Kingma2014-qo} learning rate optimizer.
(*) Although Choi et al. \cite{choi2017} report they use a learning rate of 1e-2, we only found acceptable results with a learning rate of 1e-3.
We train each network for 30 epochs.
To compensate for class imbalance, we use a random sampling technique that always creates batches with input programs and accompanying query lines such that the number of query lines with each label are equal.

\section{Evaluation}

In this study, we investigate the effectiveness of memory networks and static analyzers at predicting security defects in synthetic code through the following research questions:
\begin{itemize}
\item {\bf RQ1:} How accurate are static analyzers and memory
  networks on predicting buffer overflows as a function of code complexity?
\item {\bf RQ2:} How does the performance of static analyzers compare with
memory networks as a function of training set size?
\item {\bf RQ3:} What considerations are unique to memory networks as compared
to other tools that predict buffer overflows?
\end{itemize}

\subsection{Methods}\label{sec_methods}

The s-bAbI dataset contains four different labels for lines with buffer writes: two kinds of unsafe writes, and two kinds of safe writes, as described above.
We trained our memory networks with these four labels, so that e.g. if the network predicts that a label is \verb|BUFWRITE_TAUT_SAFE| when the true label is \verb|BUFWRITE_COND_SAFE|, the loss function penalizes this as a wrong answer.
In contrast, static analyzers give warnings when they encounter lines that they find unsafe, and are not designed to communicate any reasoning process related to control flow.
We will say that a static analyzer gives a positive result on a buffer write line whenever it generates a buffer overflow warning; we say that the tool gives a negative result when it does not issue a buffer overflow warning on that line.
Research Questions 1 and 2 are about comparing memory networks to static analyzers.
To fairly compare memory networks to static analyzers, for these questions, we collapse both ground-truth labels and network predictions to two classes, unsafe (positive) and safe (negative).
Research Question 3 is only about memory networks.
In that discussion, we use the four classes without collapsing.

To generate the s-bAbI dataset, we use a maximum of ten unique variable names of the form $\verb|entity_|n$.
We generate integers for indices and buffer lengths in the set $\{0, \dots, 99\}$.
We generate each buffer write by writing a single character from the set of digits and ASCII letters.

In our memory network, we use $H = 3$ processing ``hops.''
We use the embedding dimension $d = 32$.

As described in Section \ref{sec_labels}, we intentionally create labels under the assumption that the program does not crash when it attempts to perform an unsafe buffer write.
However, some static analyzers have a sound model of software that assumes the program stops executing after an unsafe buffer write attempt.
For these tools, we created a ``sound subset'' of the test dataset, attempting to minimize the number of buffer writes queried that occur in program execution after the first unsafe buffer write.
Since we interpret the tool's output as ``safe'' whenever it does not issue a warning, this minimizes the number of times where we inappropriately interpret ``the program never gets there'' as the tool reporting that this line is safe.

We created the sound subset by querying only the unsafe buffer write with the smallest line number, and all (safe) buffer writes with smaller line numbers, in each function.
From the set of all 76549 buffer writes in the full test set, the sound subset retains 51468 (67\%). 

This is not a perfect solution, as sometimes the first unsafe buffer write reached in a program's control flow is not the same as the unsafe buffer write with the lowest line number.
For example, if the condition of an \verb|if| statement is false, then an unsafe buffer write in the ``true'' branch will never execute, even if it is the vulnerability with the lowest line number.
However, we believe that this solution provides a fair enough testing ground for the sound tools.
Although soundness is not the main focus of this paper, our sound subset appropriately highlights the high performance of sound tools in the contexts they were designed for.

\subsection{RQ1: Performance as a function of code complexity}

In Table \ref{tab_tool_f1s}, we compare the performance of the memory network and several static analysis tools (open-source tools \verb|clang_sa|, \verb|cppcheck|, \verb|frama-c|, and a commercial tool) on the labeled buffer overflow datasets discussed.
We use the F1 metric, the harmonic mean of precision and recall, as a single-number estimate of the model's predictive performance (on a scale from 0 to 1 where higher is better):
$$
F_1 = \frac{\text{precision} \cdot \text{recall}}{\text{precision} + \text{recall}}
$$
The F1 captures the tradeoff between precision and recall, since F1 will only be high if both precision and recall are high.
The memory network's F1 scores are listed in the minimum, median, and maximum of 10 independent runs, differing only by randomness during training in the initialization of network parameters and in the random order in which training examples are shown to the network.

\begin{table}[htbp]
\caption{F1 (\%) per task. Asterisk indicates sound tool.}
\begin{center}
\begin{tabular}{|c|c|c|c|c|}
tool                    & CJOC-bAbI       & s-bAbI             & s-bAbI (sound) & Juliet \\
\hline
Mem Net           & 59.6,71.3,77.9 & 90.9,91.7,92.5 & 89.7,90.6,91.7  &   --- \\
\verb|clang_sa| & ---                      & 72.9                 &      84.9             &   8.0   \\
comm. tool        & ---                      & 93.0                &      91.9             &   40.6 \\
\verb|cppcheck| & ---                     & 80.4                 &      76.9            &   2.2    \\
\verb|frama-c|*    & ---                     & 85.2                 &      98.6             &  16.3   \\
\end{tabular}
\label{tab_tool_f1s}
\end{center}
\end{table}

\subsubsection{CJOC-bAbI}

The CJOC-bAbI column shows our attempt to replicate the results of
Choi et al. \cite{choi2017} on their own dataset \cite{CJOC-bAbI:URL}.
Specifically, we use their training set of 10,000 functions and the ``level 4'' testing set of 1,000 functions, representing the most challenging test cases.
The memory network achieves moderately good performance on the CJOC-bAbI dataset, with a median F1 of 71.3\%.
However, this is significantly less than the F1 of 82\% that they report in their paper.
Since they ``averaged the scored of the ten best cases with the smallest training error,'' but do not specify how many cases they chose the ten best out of, we believe that they may have generated many more than ten experiments, artificially inflating their performance.

The static analyzers' performance is not displayed for the CJOC-bAbI dataset because although their dataset contains ``C-like'' code, there are several issues present that prevent any example from being valid C.
Therefore, it is not possible to run static analyzers on the dataset without significantly altering the data.

\subsubsection{s-bAbI}\label{sec_methods_sbabi}

We display performance of the memory network and each static analyzer on the full s-bAbI testing set, as well as on the sound subset.
The full testing set has 38,400 files and 76,549 buffer writes, while the sound subset has 51,468 buffer writes.
The tools are able to run without altering the data because every s-bAbI example is valid C.
In this section, memory network results are shown for the largest training set, with 153,600 example files, and a total of 307,650 labeled buffer writes among all the files.
Results that evaluate the memory network's performance as a function of training set size are shown in section \ref{sec_train_size}.

Although the memory network has high performance on the s-bAbI datasets, with median F1 of 91.7\% and 90.6\%, respectively, it does not decisively outperform all of the static analyzers on either dataset.
One strength of the memory network is that its recall is higher than that of all static analyzers on the full testing set, and on all but \verb|frama-c| in the sound subset, indicating that it is able to correctly warn on a greater proportion of the unsafe lines, as seen in Table \ref{tab_recall}.
Since the memory network functions by finding patterns that look like buffer overflows, it has the opportunity to pick up on many more {\em potential} buffer overflows than static analyzers, which may fail to find a proof that a given line contains a buffer overflow.
This intuition is supported by the fact that \verb|frama-c|, a sound tool, has much greater recall than the memory network on the sound subset, where it is able to find proofs of buffer overflow a significantly greater proportion of the time.
We attribute the fact that \verb|frama-c|'s recall is less than 100\% to the imperfections of the sound subset, as described in \ref{sec_labels}, and believe that \verb|frama-c| in fact achieves true soundness on the sound subset.
This implies that the abstract interpretation model that \verb|frama-c| uses is sufficient to reason about buffer overflows on this subset.

\begin{table}[htbp]
\caption{Soundness, measured by recall (\%) per task. Asterisk indicates sound tool.}
\begin{center}
\begin{tabular}{|c|c|c|}
tool                    & s-bAbI             & s-bAbI (sound) \\
\hline
Mem Net          & 86.7, 88.2, 90.6 & 84.5, 86.0, 88.7   \\
\verb|clang_sa|  & 57.3                 &      73.9             \\
comm. tool        & 86.8                &      85.0            \\
\verb|cppcheck| & 68.8                 &      64.2            \\
\verb|frama-c|*    & 74.2                 &      97.2            \\
\end{tabular}
\label{tab_recall}
\end{center}
\end{table}

However, the pattern-finding function of the memory network causes it to also generate too many false positives, so its precision is less than that of the static analyzers, as seen in Table \ref{tab_precision}.

\begin{table}[htbp]
\caption{Completeness, measured by precision (\%) per task. Asterisk indicates sound tool.}
\begin{center}
\begin{tabular}{|c|c|c|}
tool                    & s-bAbI             & s-bAbI (sound) \\
\hline
Mem Net          & 93.9, 95.4, 96.0  & 94.0, 95.5, 96.4   \\
\verb|clang_sa|  &  99.9        &     99.9        \\
comm. tool        &    100        &     100   \\
\verb|cppcheck| &    96.7    &        95.9     \\
\verb|frama-c|*    &    100      &      100     \\
\end{tabular}
\label{tab_precision}
\end{center}
\end{table}

The high precision values for the static analyzers appear to indicate that their analysis is complete, even the sound tool \verb|frama-c|.
This is because the s-bAbI dataset does not contain safe examples that are challenging enough for these tools to mistakenly classify as unsafe.
Although we believe that these static analyzers are not complete, s-bAbI's code is not complex enough to provide evidence to prove their incompleteness.

We emphasize that since the memory network was trained only on our synthetic training dataset, its predictive capability is only high on a dataset with a similar distribution.
The testing dataset was generated from the same dataset as the training distribution, so the good performance meets expectations.
We would not expect the memory network trained on this dataset to be able to predict buffer overflows in any code that does not come from the s-bAbI data generator.

The static analyzers show fairly good performance on the s-bAbI datasets, although surprisingly not as high as we would hope, since these are fairly simple test cases without the complexity of function calls, and with a maximum of three control flow nodes.
Moreover, there is no clear best static analyzer, as shown by the difference in performance between the full s-bAbI test set and the sound subset.
For example, since \verb|frama-c| uses a sound model of program execution, it performs significantly better on the sound subset.

\subsubsection{Juliet}
The memory networks failed to converge in training on the Juliet buffer overflow dataset.
In every training run, after only a few epochs, the network's weights would be incorrectly tuned so that it would either always predict ``safe'' or always predict ``unsafe.''
This is because the Juliet dataset is too small and too complex for the memory network architecture to succeed in learning.
Although the Juliet dataset is, like CJOC-bAbI and s-bAbI, a synthetic dataset, it contains a relatively small number (5,906) of C files, containing 4,096 lines with buffer overflow.
The Juliet dataset is also complex: there are 116 unique functional variants of buffer overflow vulnerabilities presented.
Therefore for each functional variant of buffer overflow vulnerability presented, there are only a small number of files that show examples of that variety.
There is not enough data present for the memory network to successfully learn the distribution of the dataset.

We also see that the static analyzers have very low performance on the Juliet dataset, confirming previous findings \cite{Wagner2005-pp,Emanuelsson2008-hu,Johnson2013-lc,Goseva-Popstojanova2015-jg}.

\subsubsection{Discussion}
Although it appears that memory networks are competitive with static analyzers on the s-bAbI dataset, and have the advantage of greater recall, they achieve this success only when being trained on a large, simple, synthetic code dataset.
A weakness of this method is that these models are likely overfitting to this dataset, memorizing the structure of s-bAbI instead of the structure of C, as explored more in RQ2 and RQ3.
This method does not easily generalize when applied to a small, complex dataset like Juliet, as there is simply not enough data to allow the memory network to learn any of the code's relevant structure.
Although this work shows that a deep learning method has the ability to learn structure on code with non-trivial control flow, advancing the initial work by Choi et al. \cite{choi2017}, significant effort is needed to find a method that will generate to code of real-world complexity.

This work also confirms previous findings that static analyzers have low performance at predicting security vulnerabilities in the Juliet dataset.
A developer wishing to improve the performance of a static analyzer might be overwhelmed by the complexity of the Juliet dataset and the number of false positives and false negatives generated by their tool.
The s-bAbI dataset is considerably less complex than Juliet, but is still valid C code with control flow structures found in real-world code, showing places for these tools to be improved.

\subsection{RQ2: Effects of training set size} \label{sec_train_size}

\begin{figure}[h]
\includegraphics[width=8cm]{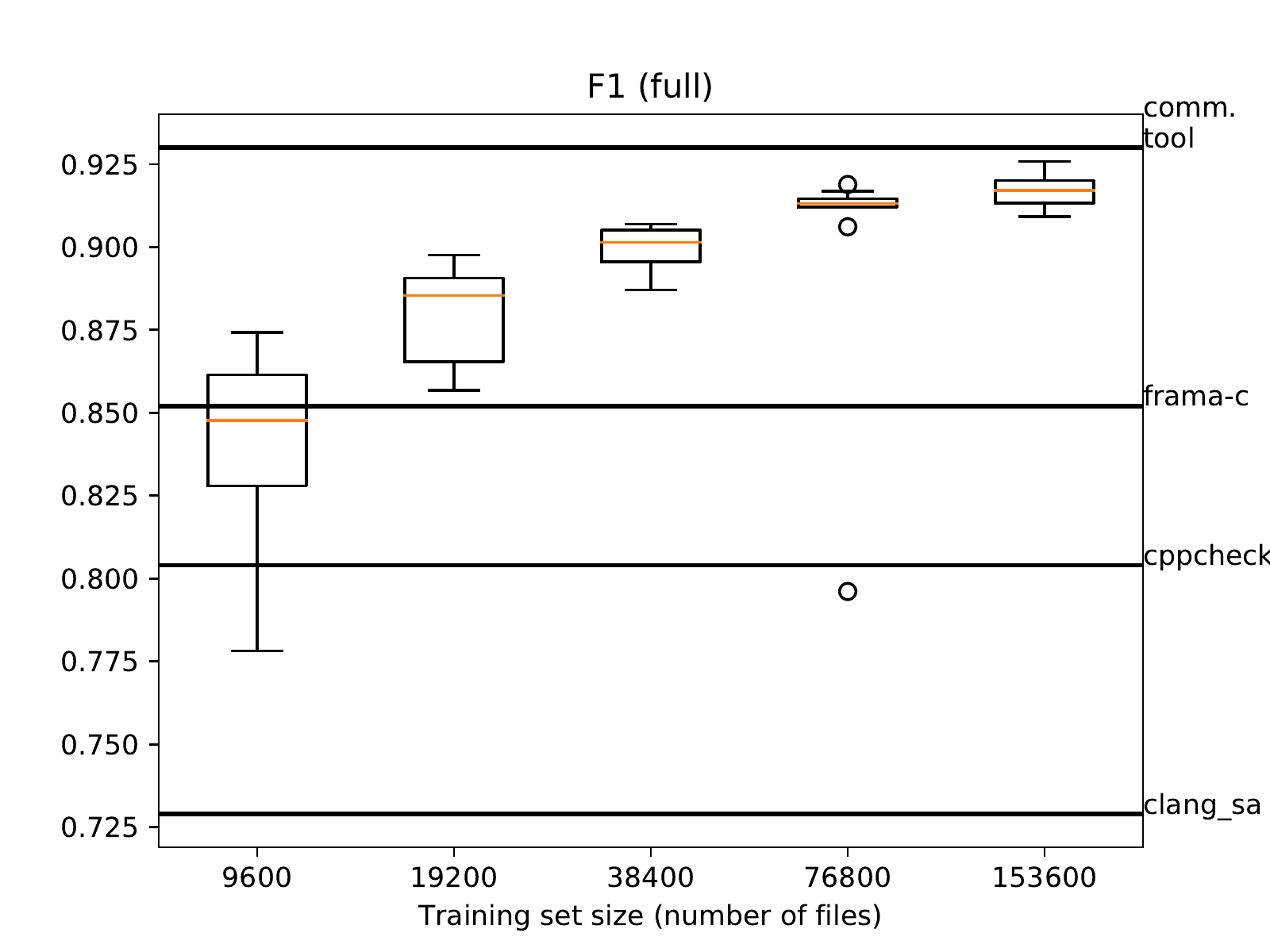}
\includegraphics[width=8cm]{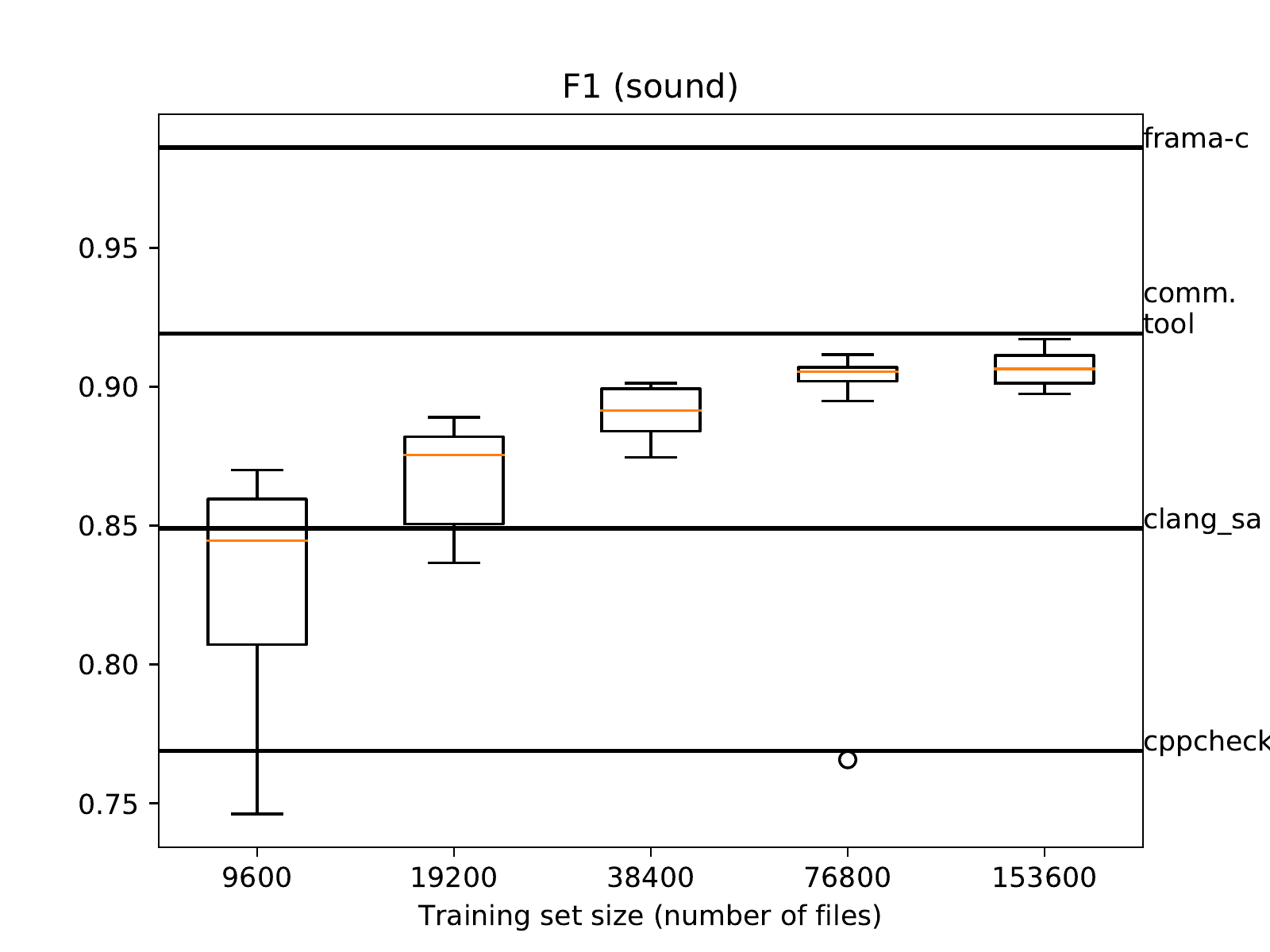}
\caption{F1-score as a function of training set size
(top) as evaluated on the entire validation subset
(bottom) as evaluated on the sound subset.}
\label{fig_f1_trainsize}
\end{figure}

For memory networks to achieve the high level of performance at predicting buffer overflow vulnerabilities in the s-bAbI dataset, as described in Section \ref{sec_methods}, they need to be trained on a large dataset, with 153,600 example files, and a total of 307,650 labeled buffer writes among all the files.
For a data-driven technique to be applied to real code, access to such a large labeled code dataset is a significant limiting factor.
To show the effect of training set size on network performance, we ran training runs with identical network parameters on training datasets of increasing size, evaluating each network on the same full testing set and sound subset as in Section \ref{sec_methods_sbabi}.
The training set size ranges from 9,600 to 153,600 files, doubling in each experiment.
We ran ten training experiments for each training set size.
The results are shown in Figure \ref{fig_f1_trainsize}.

Larger training set sizes are correlated with greater overall performance, shown by the median F1 rising with each increase of the training set size, although we see diminishing gains as the training set gets larger.
On the full validation set, the median F1s were 84.8\% for the smallest training set, followed by 88.5\%, 90.1\%, 91.3\%, and finally 91.7\% for the largest training set.

Larger training set sizes are also correlated with lower variability between runs.
The networks trained on the smallest training set are highly variable in predictive performance, with minimum and maximum F1 scores on the full testing set of 77.8\% and 87.4\%, respectively.
Increasing the training set size tends to reduce the variability, with the networks trained on the largest training set having minimum and maximum F1 of 90.9\% and 92.5\%, respectively.
We caution that there are some exceptions; the networks trained on the training set of size 76,800 mostly had very small variance in F1, in the range 90.6\%--91.9\%, but one run had an F1 of 79.6\%.
We intentionally do not exclude this as an outlier, to show that even with very large training sets, unfavorable gradient descent conditions can lead to poor performance.

Memory networks trained in the purely-supervised method of this work need a large amount of labeled training data to show competitive performance with static analyzers.
Since this amount of labeled training data about security vulnerabilities is difficult to find and expensive to create for real-world code, this simple approach is not a viable method for finding security vulnerabilities with the data sources currently available.
A training method that is more conservative with the need for labeled
data, such as a semi-supervised or active learning method, as well as
a data representation that more accurately represents the structure of
source code, such as a bag of AST paths \cite{Alon2018-mv}, or both, will be necessary to create a viable data-driven method of security vulnerability detection.

\subsection{RQ3: Properties unique to memory networks}

Here we consider some properties of the trained memory networks that help illuminate how they work differently than static analyzers.
These tests are conducted on the full s-bAbI testing set, since the memory networks do not use a sound model of program behavior and do not need to be restricted to the sound subset.
As described in Section \ref{sec_methods}, we trained our networks on four labels, which capture not only the safety of a line (\verb|SAFE| or \verb|UNSAFE|), but also the scope of reasoning required to prove the safety (\verb|COND| or \verb|TAUT|).
For the purpose of these tests, since we are no longer comparing to static analyzers, we keep all four labels as they are, without collapsing.

\subsubsection{Recognizing vulnerability type}
In addition to distinguishing safe from unsafe, the memory networks are able to distinguish conditional buffer writes from tautological, as shown in the confusion matrix in Table \ref{tab_confusion}.
Across the ten networks that were trained on the largest training set, we took the median value of each cell in their confusion matrices.
Rows indicate the true label. Columns indicate the predicted label.

\begin{table}[htbp]
\caption{Median confusion matrix.}
\begin{center}
\begin{tabular}{c|cccc}
                    & \verb|C_S| & \verb|C_U| & \verb|T_S| & \verb|T_U| \\
\hline
\verb|C_S|   &     11,378.5   &     1324.5    &        0       &         0         \\
\verb|C_U|   &       4,230      &     21,465.5 &         0        &       0          \\
\verb|T_S|    &        0        &          0        &     18,226    &      672      \\
\verb|T_U|   &        0         &          0         &      947     &      18,303      \\
\end{tabular}
\label{tab_confusion}
\end{center}
\end{table}

A block-diagonal structure is evident, showing that the memory network is decisively able to distinguish between conditional and tautological buffer writes.
Since the integer indices for conditional buffer writes always appear in the conditions of the \verb|if| statements and the loop guards of the \verb|for| and \verb|while| loops, and the indices for the tautological buffer writes never appear in these places, we believe that the network is making the conditional-or-not decision based on the co-occurrence of variable names in lines with these control flow keywords.

\subsubsection{Failure to generalize}

The memory networks are trained on a corpus with a fixed vocabulary, namely, the set of all unique tokens in the s-bAbI dataset.
Their predictive performance, while strong on the testing set that has the same set of tokens as the training set, does not generalize to a codebase that contains any tokens not contained in the training set.
Since each token in a given data instance passes through the embedding matrices $E_\text{val}$ and $E_\text{addr}$, which contain one row for each element of the training-set vocabulary, any novel token must be mapped to one of the tokens seen in the training set for computation to continue, but this mapping necessarily destroys information crucial to the network's performance.

For example, the training set contains tokens for the integers 0 through 99.
Suppose that we had a modified testing set, generated by converting every integer literal $n$ into $n + 1000$.
The labels on each line would still be correct, because the integer indexes and the array lengths are all incremented by the same amount.
However, none of the integer literals 1000 through 1099 are in the training set, so these novel tokens need to be remapped.
If we were unaware that this was the transformation that had been performed, we would not have any information about an informative remapping, so it would be natural to choose an uninformative one, namely mapping all integer literal tokens to a single integer literal token in the vocabulary.

We ran this experiment by mapping all tokens representing integer literals to the token representing 0 in the test set, then running the same ten trained memory networks on this modified dataset.
Table \ref{tab_unseen_confusion} shows the resulting confusion matrix.

\begin{table}[htbp]
\caption{Median confusion matrix on unseen integer set.}
\begin{center}
\begin{tabular}{c|cccc}
                    & \verb|C_S| & \verb|C_U| & \verb|T_S| & \verb|T_U| \\
\hline
\verb|C_S|   &     2,956      &      9,741     &        0         &         0         \\
\verb|C_U|   &     3,083      &    22,604     &         1        &       0          \\
\verb|T_S|    &        0        &          0        &     16,859    &      2,038    \\
\verb|T_U|   &        0         &          1         &     17,066   &      2,182     \\
\end{tabular}
\label{tab_unseen_confusion}
\end{center}
\end{table}

We see that the block-diagonal structure still exists, since it is still possible to identify whether a buffer write's safety is conditional or tautological by seeing if its index co-occurs with control-flow-relevant tokens.
But within the blocks, we see a notable decrease in the memory network's ability to distinguish safe from unsafe.

The set of integer literal tokens in C is large, but finite.
Even so, it is unlikely that any code dataset used for training, no matter how large, will contain all of the integer literal tokens that could show up in a query to a trained model.
Furthermore, the set of string literals, and the set of valid variable names, are both infinite.
No code dataset will contain all possible such tokens.
Models with the ability to gracefully digest novel tokens, or a redefinition of tokens that ensures that all possible tokens are seen during training time, are needed before AI on code is a real possibility.
For example, representing numerical tokens as numerical quantities, such as in Neural Arithmetic Logic Units \cite{Trask2018-nm}, will be necessary for success on fine-grained code-understanding tasks.





\section{Related Work} \label{sec_related_work}


Our works builds on three distinct threads in the literature: research
on question answering tasks, the emerging literature of artificial
intelligence for software engineering, and the development of data
sets to assess software security tools.

\subsection{Question Answering Tasks}

Question answering is a classic problem in text retrieval. For
example, the NIST sponsored Text REterival Conference (TREC) has had a
Question Answering track since 1999 \cite{Voorhees2000-fx}. Current
question answering systems are able to rival human performance for a
subset of tasks focusing on answering realistic reading comprehension
questions \cite{squad-zk,Yu2018-bi}.

Our work is closely related to that of Weston's group at Facebook,
where they took an alternative approach: for a simple neural network,
find the limits of the questions it can answer by generating
synthetic, labeled data which are sufficiently complex to break the
network, and then improve the neural network until the tasks are able
to be cleared \cite{weston2014,Weston2015-tu}. More specifically, the
memory network architecture first proposed by Weston et
al. \cite{weston2014} to answer reading comprehension questions about
short stories in the English language, trained and tested on the
original bAbI dataset \cite{Weston2015-tu}. Weston et al's training
scheme involved supplying the network with a story, broken into
sentences, a single-sentence question about the story.  The
ground-truth data that they used during training included both a
single-word correct answer to the question, and the set of sentences
in the story that were relevant to answer the question.  Sukhbaatar et
al. \cite{sukhbaatar2015} refined the memory network architecture so
that the set of relevant sentences did not need to be supplied during
training, an improvement that they give the name {\em end-to-end
  training}. We use this approach in our work, although in the domain
of source code instead of natural language. 

\subsection{Artificial Intelligence for Software Engineering}

There is much recent interest in applying artificial intelligence and
machine learning techniques to a variety of software engineering
tasks; see \cite{Allamanis2017-wv} for a comprehensive
review. The two papers closest to our work are Russel et
al. \cite{Russell2018-fx}, which attempts to identify vulnerabilities
at the function level, and Choi et al. \cite{choi2017}. which attempts
to identify vulnerabilities at the line level. 

Our work builds on Choi et al. \cite{choi2017}, which used a variant
of the Sukhbaatar et al. \cite{sukhbaatar2015} architecture to predict
buffer overflow vulnerabilities at the line level in synthetic source
code.  They trained and tested on a synthetic code dataset that they
created, which we refer to as {\em CJOC-bAbI}. Their work, however, has
several limitations: first, CJOC-bAbI was not valid C, making it
difficult to compare against existing static analysis tools. Second,
CJOC-bAbI only generated basic blocks, and the absence of loops,
conditionals, and variables of unknown value make their test cases far
too simple.
 
Our work is a natural extension of Choi et al. \cite{choi2017} where
we address the major shortcomings of their paper: we generate valid C
code instead of `C-like' code; we incrementally increase the
complexity of the code generated to include conditionals and loops;
and we study how well the combination of the representation used to
input the code into the neural network and the network architecture
interact during training at the line level. 

\subsection{Software Security Data Sets}

The development and testing of software security data sets is well
established in the literature. We note, however, that much of the
existing work on software security data sets focuses primarily on
finding realistic defects in realistic code, a high, and important
threshold to cross. For example, the NIST Software Assurance Metrics
And Tool Evaluation (SAMATE) project \cite{Black2006-ho}, which has
coordinated the release of several data sets (Juliet
\cite{Black2018-le}, SARD \cite{Black2018-ae}, IARPA Stone Soup \cite
{noauthor_undated-cm} and hosted competitions for static analysis
tools \cite{Okun2009-iz, Okun2010-zv, Okun2011-at,Okun2013-ru},
focuses on realistic test cases.  Similarly, LAVA
\cite{Dolan-Gavitt2016-zg} and many other efforts
(\cite{Zitser2004-ep,Ayewah2007-ra,Wagner2005-pp,Shiraishi2015-vg,Kratkiewicz2005-rg,Pashchenko2017-xn})
have focused on the realism of the defect in realistic code settings.

Although the goal of these projects is to find realistic defects in
realistic code, their performance is not yet high enough; i.e. Table
\ref{tab_tool_f1s} and De Oliveira et al. \cite{De_Oliveira2017-xm}.

Our work, in contrast, is more modest: we attempt to find the minimal
code complexity necessary to break the state of the art AI system
\cite{choi2017}. In doing so, we find that a memory network can learn
how to identify buffer overflows in synthetic code, but the it appears
as though the memory network needs a nearly exhaustive amount of
training data in order to do so.

\section{Conclusions and Future Work}
In this study, we investigate the limits of the current state of the
art AI system for detecting buffer overflows and compare it with
current static analysis engines. To do so, we developed a code
generator, {\em s-bAbI}, capable of producing an arbitrarily large number of
samples of controlled complexity. We found that the static
analysis engines we examined have good precision, but poor recall on
this dataset,
except for a sound static analyzer that has good precision and recall. We
found that the state of the art AI system, a memory network modeled
after Choi et al. \cite{choi2017}, can achieve similar performance to
the static analysis engines, but requires an exhaustive amount of
training data in order to do so. 

Our work implies that there are three threads of future work: First,
further developing static analysis engines to improve their recall
against this minimally complex class of synthetic code as a lower bar
than NIST's more realistic code datasets. Second, improving AI
systems to the point were they can at least solve {\em s-bAbI}. And,
third, increasing the complexity of {\em s-bAbI} to find the
additional failure modes of improved static analysis engines and AI
systems.

Future work on the latter two threads might include improving the
representations of how code is encoded for the AI system (e.g. directly
modeling scope, such as with a bag of AST paths \cite{Alon2018-mv}),
as we suggest in RQ2, and
modifying the network to be able to compute on integer representations
(e.g. integrating Neural Arithmetic Logic Units \cite{Trask2018-nm})
instead of merely memorizing the finite amount of training data and
relying on inner product search), as we suggest in RQ3.
The performance characteristics of
these modeling approaches would then give insight into how to modify
{\em s-bAbI} so that the new code would defeat the new models. By
iterating in this way, we are hopeful to improve the state of security
defect prediction.

\section{Acknowledgements}
\noindent Copyright 2018 Carnegie Mellon University. All Rights
Reserved.  
~\\[1em]
\noindent This material is based upon work funded and supported by the
Department of Defense under Contract No. FA8702-15-D-0002 with
Carnegie Mellon University for the operation of the Software
Engineering Institute, a federally funded research and development
center. 
~\\[1em]
\noindent The view, opinions, and/or findings contained in this material are
those of the author(s) and should not be construed as an official
Government position, policy, or decision, unless designated by other
documentation.  
~\\[1em]
\noindent References herein to any specific commercial product, process, or
service by trade name, trade mark, manufacturer, or otherwise, does
not necessarily constitute or imply its endorsement, recommendation,
or favoring by Carnegie Mellon University or its Software Engineering
Institute.  
~\\[1em]
\noindent NO WARRANTY. THIS CARNEGIE MELLON UNIVERSITY AND SOFTWARE ENGINEERING
INSTITUTE MATERIAL IS FURNISHED ON AN "AS-IS" BASIS. CARNEGIE MELLON
UNIVERSITY MAKES NO WARRANTIES OF ANY KIND, EITHER EXPRESSED OR
IMPLIED, AS TO ANY MATTER INCLUDING, BUT NOT LIMITED TO, WARRANTY OF
FITNESS FOR PURPOSE OR MERCHANTABILITY, EXCLUSIVITY, OR RESULTS
OBTAINED FROM USE OF THE MATERIAL. CARNEGIE MELLON UNIVERSITY DOES NOT
MAKE ANY WARRANTY OF ANY KIND WITH RESPECT TO FREEDOM FROM PATENT,
TRADEMARK, OR COPYRIGHT INFRINGEMENT.  
~\\[1em]
\noindent [DISTRIBUTION STATEMENT A] This material has been approved for public
release and unlimited distribution.  Please see Copyright notice for
non-US Government use and distribution.
~\\[1em]
\noindent This work is licensed under a Creative Commons Attribution-NonCommercial-NoDerivatives 4.0 International License.
~\\[1em]
\noindent Carnegie Mellon$^\text{\textregistered}$ and CERT$^\text{\textregistered}$  are registered in the U.S. Patent and Trademark Office by Carnegie Mellon University.
~\\[1em]
DM18-0988	\hfill \href{http://creativecommons.org/licenses/by-nc-nd/4.0/}{\includegraphics[height=3em]{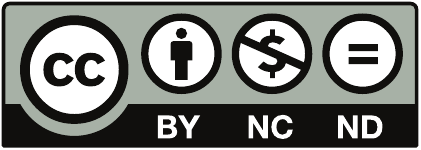}}	


\bibliography{icse-references}

\end{document}